\documentclass[%
 aip,
 jmp,%
 amsmath,amssymb,
 reprint,%
]{revtex4-1}

\usepackage{graphicx}
\usepackage{xcolor}
\usepackage{subfigure}
\usepackage{aas_macros}
\usepackage{natbib}
\usepackage{dcolumn}
\usepackage{bm}
\usepackage[nolist]{acronym}

\begin{document}


\title{A Double Vacuum Window Mechanism for Space-borne Applications}

\author{K. Zilic}
 \affiliation{University of Minnesota/Twin Cities, School of Physics and Astronomy, 116 Church St, Minneapolis, MN 55455, USA}

\author{A. Aboobaker}
\affiliation{Jet Propulsion Laboratory, 4800 Oak Grove Dr, Pasadena, CA 91011, USA}

\author{F. Aubin}
 \affiliation{University of Minnesota/Twin Cities, School of Physics and Astronomy, 116 Church St, Minneapolis, MN 55455, USA}

\author{C. Geach}
 \affiliation{University of Minnesota/Twin Cities, School of Physics and Astronomy, 116 Church St, Minneapolis, MN 55455, USA}

\author{S. Hanany}
 \affiliation{University of Minnesota/Twin Cities, School of Physics and Astronomy, 116 Church St, Minneapolis, MN 55455, USA}
\email{hanany@physics.umn.edu}

\author{N. Jarosik}
\affiliation{Department of Physics, Princeton University, Princeton, NJ 08544, USA}

\author{M. Milligan}
 \affiliation{University of Minnesota/Twin Cities, School of Physics and Astronomy, 116 Church St, Minneapolis, MN 55455, USA}

\author{I. Sagiv}
 \affiliation{University of Minnesota/Twin Cities, School of Physics and Astronomy, 116 Church St, Minneapolis, MN 55455, USA}

\date{\today}

\begin{abstract}

We present a vacuum window mechanism that is useful for applications requiring two different vacuum windows in series, 
with one of them movable and resealable. 
Such applications include space borne instruments that can benefit from a thin vacuum window at low ambient pressures, 
but must also have an optically open aperture at atmospheric pressures. We describe the implementation and 
successful operation with the EBEX balloon-borne payload, a millimeter-wave instrument designed to measure 
the polarization of the cosmic microwave background radiation.


%
\end{abstract}

\maketitle

%

\section{Introduction}
\label{sec:introduction}

The use of vacuum windows is ubiquitous throughout science and engineering. In certain applications there is a 
need to implement two vacuum windows in series. Examples include balloon-borne or space-based applications in 
which one prefers to use a thicker window for the larger differential pressure on the ground, and a thinner
window when the payload is above Earth's atmosphere. If the experiment is to be reusable, 
the vacuum window mechanism must be reversible, that is, the atmospheric pressure seal must be resealable. 

Ground testing and calibration of experiments that are sensitive to electromagnetic radiation require that 
both vacuum windows in the series be transparent to the desired wavelengths. The choice of vacuum window 
material depends, among other factors, on the wavelength of use. Many such 
applications require windows with low loss, low reflection, and low emission. Useful 
materials in the millimeter wavelength include polyethylene and polypropylene. They are readily 
available in a large range of thicknesses and sizes, and have low cost. They have 
relatively low index of refraction, making the fabrication of anti-reflection coating less of a challenge compared 
to higher index materials. Ultra high molecular weight polyethylene (UHMWPE) and polypropylene have loss tangents 
of  $8\times10^{-5}$ and $4\times 10^{-4}$, respectively, 
among the lowest at this waveband~\cite{Cardiffabsorption, lamb96}.

In this paper we describe a vacuum window that we developed for the balloon-borne E and B Experiment (EBEX). 
It consisted of two polyethylene vacuum windows in series, one of them removable and 
resealable. We call the apparatus the \ac{DWM}. In Section~\ref{sec:ebex} we 
describe the requirements that led to the development of the \ac{DWM}; and in Section~\ref{sec:dwm_design} we discuss the 
specifics of the design of the \ac{DWM} as well as the testing performed on the \ac{DWM} and the results found.


\section{EBEX}
\label{sec:ebex}


\ac{EBEX} was a stratospheric balloon-borne experiment designed to measure the polarization of the cosmic 
microwave background radiation~\cite{ebexpaper1,ebexpaper3}.
It consisted of an ambient temperature telescope focusing light 
into a cryogenic receiver. The receiver had an array of nearly a thousand bolometric transition edge 
sensors operating at a temperature of 0.25~K. The experiment had three frequency bands centered 
on 150, 250 and 410~GHz and collected data in a flight circumnavigating Antarctica in January 2013. 

The optical design determined the 300~mm open diameter of the receiver's vacuum window. 
Below the window we placed reflective filters to reject high frequency radiation. Space constraints dictated that 
these filters be placed no farther than 10~mm below the vacuum window, limiting the maximum bowing acceptable 
for the window material before damaging the filters. 

We considered several materials for the vacuum window including alumina, silicon, sapphire, Zotefoam, polypropylene, and 
varieties of polyethylene. We chose \ac{UHMWPE} because it has low loss, a relatively low index and our collaborators 
had already developed a broad-band anti-reflection coating for it; because it is not fragile; and because it 
is readily available at many sizes and thicknesses. 

We measured the central deflection of 300~mm diameter \ac{UHMWPE} window under atmospheric differential 
pressure for a number of thicknesses. We found that a minimum of 12.7~mm thick material was needed to provide
a deflection of less than 10~mm.  Although total absorption with this thickness is only 0.5\%, 1.1\%, and 3.0\%, at the 
three EBEX frequency bands, the emission of such 
a room temperature window would have represented 8\%, 11\%, and 14\% of the total optical load absorbed 
by the detectors at 150, 250, and 410~GHz, respectively. The temperature at float altitude is 
close to room temperature. To make the optical load and resultant photon noise from the window negligible compared 
to other sources, we decided to use two windows in series: a removable thick window for ground operations, and, below it,
a thinner window only for float altitude. A comparison between the optical load for the two thicknesses 
is given in Table~\ref{tab:loading}. We use the words `above' and `below' to indicate 
relative positions closer to the higher and lower pressures, respectively. 

\begin{table}[h]
\begin{tabular}{| c | c | c |}
\hline
Band  & 12.7 mm window & 1 mm window \\
\hline 
150 GHz & 8 \% & 0.1 \%\\
\hline 
250 GHz & 11 \% & 1 \%\\
\hline
410 GHz & 14 \% & 1 \%\\
\hline
\end{tabular}
\caption{ Calculated fractional in-band optical loading on the EBEX detectors 
due to a single window of either 12.7~mm or 1~mm thickness in the optical path.}
\label{tab:loading}
\end{table}

\section{Double Window Design}
\label{sec:dwm_design}

\subsection{Overview}

\begin{figure}[ht!]
  \centering
  \includegraphics[width=0.48\textwidth]{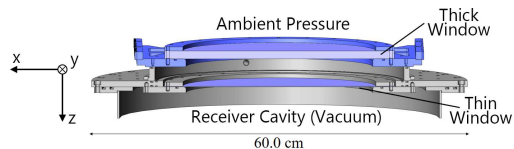}
  \caption{Cross-section view of the \ac{DWM}. The thick window was movable in the y direction, 
  exposing an open aperture above the thin window (see Figure~\ref{fig:dwm_sketch_pic}). It was also resealable. 
  The cavity between the windows was connected with a tube and valve (not shown) to the receiver cavity.  
  \label{fig:cross_section} }
\end{figure}

\begin{figure}[ht]
  \centering
 \includegraphics[width=0.48\textwidth]{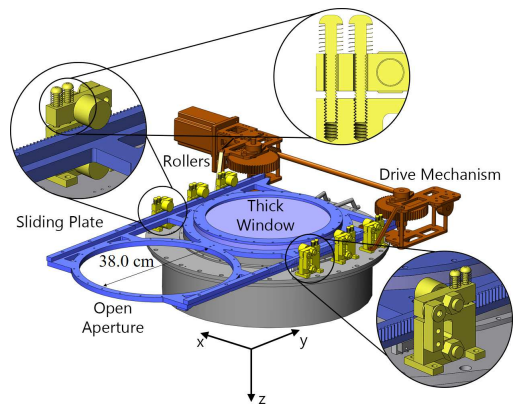}
  \caption{Model of the \ac{DWM} assembly. It consisted of a plate with two apertures (blue), a drive mechanism (brown),
   and six spring-loaded rollers, each with two springs (yellow). The mechanism gave the capability to either have
  the thick window seal the receiver, the configuration shown, or move the sliding plate in the $+y$ direction 
  and replace the thick window with an open aperture.
  \label{fig:dwm_sketch_pic} }
\end{figure}

The \ac{DWM} consisted of a permanent 1~mm \ac{UHMWPE}, anti-reflection coated window, which 
we call the `thin window'. On the ambient pressure 
side of the thin window we placed a movable plate with two apertures. One 300~mm aperture had a 12.7~mm thick 
\ac{UHMWPE}, anti-reflection coated `thick window'; the other 380~mm aperture was open. Figure~\ref{fig:cross_section} 
shows a cross-section of the construction and Figure~\ref{fig:dwm_sketch_pic} shows a solid model of the entire apparatus. 
A copper tube with a valve connected the receiver cavity to the volume immediately above the thin window. 
On the ground the thick window was always positioned above the thinner one. 
When the receiver was evacuated we also evacuated the chamber between the two 
windows. Before the payload was launched we closed the valve connecting the 
receiver and intra-window cavity.  When the payload reached float altitude we actuated a motor to move 
the two-aperture plate and place the open aperture in position above the thin window. Before flight termination 
the motor was actuated again to move the thick window back into its ground-operations position. 


\subsection{Thin Window} 

\begin{figure}[htp]
  \centering
  \includegraphics[width=0.45\textwidth]{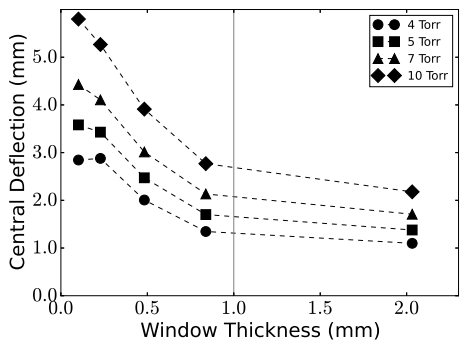}
  \caption{Central deflection versus window thickness for different pressure differentials. A 1.0 mm window was chosen.}
  \label{fig:thin_wind_defl}
\end{figure}

The 3636~kg total suspended weight below the 963,000~m$^{3}$ helium balloon gave high likelihood for 
flight altitude above 33~km and therefore an ambient pressure below $\sim$6~torr. Figure~\ref{fig:thin_wind_defl} 
shows our measurement of the central deflection 
of a 300~mm \ac{UHMWPE} window as a function of window thickness for several differential pressures of 4~torr and above. The 
differential pressures span equivalent altitudes between 29 and 35~km~\cite{standardatmosphere}. 
We chose a thickness of 1~mm because it 
gave negligible additional emission, and even at altitude as low as 28 km its maximum deflection was only 3~mm, giving 
ample space margin from the infrared filters below it. Two flat aluminum rings with 
inside (outside) diameter of 310 (368)~mm
were glued to the top and bottom of the thin window with Miller Stephenson Epoxy 907 and bolted against
an o-ring situated in a standard o-ring groove. 

 \subsection{Movable, Resealable Window} 

The \ac{DWM} mechanism had two main elements: motion and seal. (We followed 
a design similar to one initially implemented by the XPER experiment~\cite{staggs96}.)

\begin{figure}[ht]
  \centering
  \includegraphics[width=0.45\textwidth]{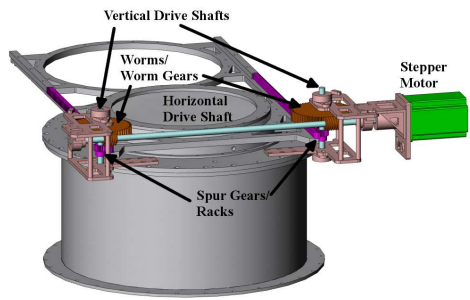}
  \caption{The \ac{DWM} drive mechanism consisted of a stepper motor (green) driving a 
  horizontal shaft drive shafts (cyan) with two worm/worm gears (brown), coupling 
  to two vertical shafts (cyan) with two spur gear/racks (magenta).
  \label{fig:dwm_drive} }
\end{figure}

The motion part of the \ac{DWM} consisted of a stepper motor driving a horizontal shaft with two worms, one on 
either side of the moving plate; see Figure~\ref{fig:dwm_drive}. The worms coupled to worm gears on two 
vertical shafts that also had spur gears. The spur gears coupled to racks mounted on either side of the moving 
plate. The plate rode between 6 pairs of 300-grade stainless steel rollers, 3 pairs on
each side of the plate. 

A gear system was used due to its simplicity, compactness, and high output torque, as well as 
tolerance for low-temperatures, such as may be experienced during the ascend period of flight with temperatures down to -50$^\circ$C. The rack was mounted
on the sides of the sliding plate, rather than on top or bottom, because the seal/reseal function 
was provided through motion in the $z$ direction (see Figures~\ref{fig:cross_section} and~\ref{fig:dwm_sketch_pic}).  
We used two symmetric spur gear and worm/worm gear systems 
on either side of the sliding plate to ensure proper movement. One of the worm/worm gear systems 
was right-handed and the other was left-handed due to the mirror symmetry of the sliding plate. 

\begin{figure}[htp]
  \centering
  \includegraphics[width=0.48\textwidth]{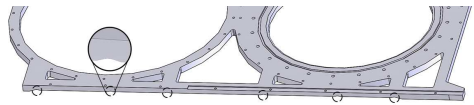}
  \caption[\ac{DWM} sliding plate]{The \ac{DWM} sliding plate with triangular notches highlighted on the bottom surface.}
  \label{fig:dwm_sl_plate}
\end{figure}

The thick window was permanently clamped directly onto a Buna-N o-ring on the movable plate. 
We also used an o-ring to facilitate the seal between the moving plate and the stationary surface
of the \ac{DWM}; both were made of aluminum. 
A relatively high hardness, Shore A~75, Buna-N o-ring was mounted onto the stationary surface. 
The movable vacuum seal required two functions, sealing in fixed known locations, and 
sliding between these locations without damaging the o-ring. Importantly, no vacuum sealing function 
was required during the motion. To achieve these functions we provided for a small $z$ displacements
of the moving plate by means of triangular notches on the bottom side of the moving plate; 
see Figure~\ref{fig:dwm_sl_plate}. The position of the slots matched the positions of the 
bottom rollers in seal positions. In these positions the effectively thinner plate would move down and 
seal against the o-ring. Sealing force was provided 
by springs that forced the top rollers in the direction of the bottom rollers. 
During motions, the plate rolled out of the notches against the force of the springs, thus moving 
away from the o-ring seal. 

The triangular notches in the sliding plate made an angle of 26.7$^\circ$ relative 
to the track face and were 3.3~mm deep. The geometry of the notches was chosen to give 
clearance for the sliding plate to travel above 
the o-ring when moving and to compress the o-ring when the 25.4~mm 
diameter rollers were seated in the notches.
The o-ring was seated in a dovetail groove which set its maximum linear compression to be 0.71 mm, 
20\% of the 3.53 mm diameter. 
This compression and the o-ring hardness required a compressive force per length of 1.2 to 
6.6~N/mm for proper vacuum sealing and o-ring durability~\cite{parkerhandbook}. We chose a value of 2.9~N/mm, 
and with the 410~mm nominal diameter o-ring, this gave a total required compressive force of 3700~N. 
In a ground configuration, a force of 13300~N was exerted uniformly on the o-ring due to the pressure differential 
across the thick window; however, at float altitude, the pressure differential drops, providing only 180~N of force. Furthermore, upon resealing at the end of flight, there is no pressure differential across the thick window at all.
To provide the required force during flight, 12 springs with spring constant of 41.3~N/mm 
were compressed 7.5~mm each by engaging the compression screws by hand, which then pressed on the 
sliding plate and o-ring via the rollers located above the sliding plate; see Figure~\ref{fig:dwm_sketch_pic}. 

We calculated that a force of 2300~N was required to move the rollers out of the notches. Given the radius of the spur 
gears of 12.7 mm and the 60:1 worm/worm 
gear ratio, this required a minimum motor torque of 0.49~Nm. However, worm/worm gear efficiency depends 
on the coefficient of friction between the two surfaces as,
\begin{equation}
  \mathrm{Efficiency} = \tan{\gamma}\frac{1 - \mu \tan{\gamma}}{\mu + \tan{\gamma}}
\end{equation}
where $\gamma=4.8^\circ$  was the worm lead angle~\cite{wormgear_eff}. We assumed that the kinetic coefficient 
of friction $\mu$ between the cast iron worm gear and steel worm was 0.2\cite{CoeffFrict} when unlubricated, which gives an 
efficiency of 29\%. In this case the minimum required torque is 1.69~Nm. 
We used Dow Corning Molykote dry lubricant to decrease the friction between the components 
and protect the open gearing from oxidation. With the dry lubricant we expected a coefficient of friction 
between 0.02 and 0.06\cite{CoeffFrict}, and thus an increase in the worm/worm gear efficiency to 
between 58\% and 81\%, and a reduction in the required motor torque to between 0.84 and 0.60~Nm, respectively. 
A stepper motor with a stall torque of 3.88 Nm (Kollmorgen model M093-LE14) was chosen to give more than a factor 
of two safety for the unlubricated case.

The \ac{DWM} was operated at low speeds, taking 3.3 minutes to move the 419~mm from the thick 
window to open aperture states. At this speed the stepper motor operated 
in its high-torque regime. Since the \ac{DWM} was operated only twice during flight, its operation 
time had negligible effect on total observation time. We monitored the position of the sliding plate 
with two electrical limit switches that were located at either end of the travel and were depressed 
when the sliding plate was in the thick window or open aperture positions. 
The state of these switches were continuously read out as analog voltages.

Aluminum components were used extensively and all steel components were
light-weighted. In total, the \ac{DWM} had a mass of $22$ kg, including the 
windows, vacuum valve, bellows, and stepper motor. 
Military specification Buna-N o-rings rated to $-54^\circ$C were used for their thermal 
properties and abrasion durability.

\section{Testing and In-Flight Operation}
\label{sec:dwm_testing}

We tested the \ac{DWM} in vacuum and flight-like temperatures in an environmental chamber at the 
Columbia Scientific Balloon Facility in Palestine, TX. Tests were conducted twice, once 
in the summer of 2011 and again in the summer of 2012. For testing purposes the \ac{DWM} was 
mounted on a fixture that simulated the receiver vacuum. We monitored the intra-window cavity pressure
to check the integrity of the thick window's vacuum seal and reseal. We placed temperature sensors on several key 
components of the \ac{DWM} including the stepper motor, sliding plate, and the plate that simulated
the receiver, and we had an ambient chamber temperature sensor. We also readout the position switches 
and chamber pressure continuously.


The simulated receiver and intra-window cavities were 
evacuated to a pressure of few torr when the apparatus was outside the environmental chamber. The valve connecting 
the two cavities was closed to separate them, as would be done pre-flight.  The \ac{DWM} testing 
apparatus was then placed in the chamber. 
The chamber was cooled to approximately $-45^\circ$ C, 
which is near the temperature experienced during initial ascent of the payload.  
The chamber and \ac{DWM} were then allowed to warm up and the following cycle was tested $6$ and $10$ 
times for the two testing periods, respectively: pump the chamber down to between $2$ and $10$ torr of 
pressure, move the sliding plate to the open aperture position, move the sliding plate to the thick window 
position, and backfill the chamber to $\sim 100$ torr with N$_2$ gas. 
The data showed robust motion of the moving plate throughout the necessary range. It also showed 
proper resealing of the intra-window cavity. 



EBEX was launched from the balloon Facility at McMurdo Station in Antarctica at 00:30 GMT 
December 29, 2012. The payload achieved an altitude of approximately 36 km about 5 hours later. 
The payload was science operational until 06:00 GMT January 9, 2013, for a duration of 10.83 days, at 
which point the liquid helium cryogen expired.  

\begin{figure}[htpb]
  \centering
  \includegraphics[width=0.48\textwidth]{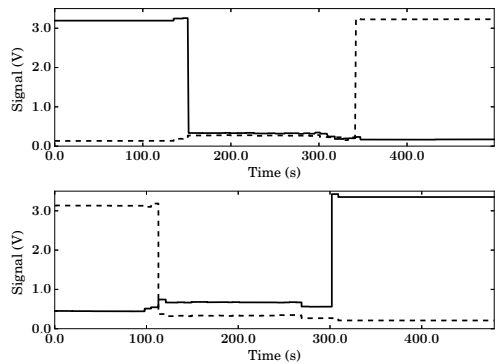}
  \caption{\ac{DWM} position encoders' signals during opening (upper) and closing (lower). High voltage 
  indicates `true' for the closed (solid) and open (dash) signals.}
  \label{fig:dwm_flight}
\end{figure}

We operated the \ac{DWM} twice during the flight.  
Three hours after launch at an altitude of 36.6 km and \ac{DWM} 
temperature of 15$^\circ$C we removed the thick window.  The second time was  
approximately two hours after liquid helium expired 
at an altitude of 36.4 km and \ac{DWM} temperature of 39$^\circ$C when we repositioned the thick 
window above the thin in preparation for flight termination.  The position encoder signals during 
these times are shown in Figure~\ref{fig:dwm_flight}. These monitors indicate proper 
opening and closing the window. Visual inspection post-flight showed the thick window in nominal 
position above the thin window. Both windows and the fragile filters below them were recovered 
intact post-flight indicating that there was no major leak by the thick window either before it was opened 
or after it was resealed. Had there been a major leak, the thin window would have ruptured, or bowed sufficiently 
to tear the filters mounted below it. Nominal cryogenic operation of the receiver through flight and the fact 
that liquid helium hold time was commensurate with pre-flight predictions indicated
nominal gas pressure inside the cryostat and therefore the absence of gas leaks through the 
thin window. We conclude that the \ac{DWM} performed successfully.

\begin{acknowledgments}

Support for the development and flight of the EBEX instrument was provided by NASA 
grants NNX12AD50G, NNX13AE49G, NNX08AG40G, 
and NNG05GE62G, and by NSF grants AST-0705134, and ANT-0944513.   
Zilic acknowledges support by the Minnesota Space Grant Consortium.  
We are grateful to Suzanne Staggs for providing the original 
XPER window upon which our design was based.   
We thank Xin Zhi Tan for help with figures.

\end{acknowledgments}

\begin{acronym}
    \acro{ACS}{attitude control system}
    \acro{ADC}{analog-to-digital converters}
    \acro{ADS}{attitude determination software}
    \acro{AHWP}{achromatic half-wave plate}
    \acro{AMC}{Advanced Motion Controls}
    \acro{ARC}{anti-reflection coating}
    \acro{ATA}{advanced technology attachment}
    \acro{BRC}{bolometer readout crates}
    \acro{BLAST}{Balloon-borne Large-Aperture Submillimeter Telescope}
    \acro{CANbus}{controller area network bus}
    \acro{CMB}{cosmic microwave background}
    \acro{CMM}{coordinate measurement machine}
    \acro{CSBF}{Columbia Scientific Balloon Facility}
    \acro{CCD}{charge coupled device}
    \acro{DAC}{digital-to-analog converters}
    \acro{DASI}{Degree~Angular~Scale~Interferometer}
    \acro{dGPS}{differential global positioning system}
    \acro{DfMUX}{digital~frequency~domain~multiplexer}
    \acro{DLFOV}{diffraction limited field of view}
    \acro{DSP}{digital signal processing}
    \acro{DWM}{double window mechanism}
    \acro{EBEX}{E~and~B~Experiment}
    \acro{EBEX2013}{EBEX2013}
    \acro{ELIS}{EBEX low inductance striplines}
    \acro{EP1}{EBEX Paper 1}
    \acro{EP2}{EBEX Paper 2}
    \acro{EP3}{EBEX Paper 3}
    \acro{ETC}{EBEX test cryostat}
    \acro{FDM}{frequency domain multiplexing}
    \acro{FPGA}{field programmable gate array}
    \acro{FCP}{flight control program}
    \acro{FOV}{field of view}
    \acro{FWHM}{full width half maximum}
    \acro{GPS}{global positioning system}
    \acro{HPE}{high-pass edge}
    \acro{HWP}{half-wave plate}
    \acro{IA}{integrated attitude}
    \acro{IP}{instrumental polarization} 
    \acro{JSON}{JavaScript Object Notation}
    \acro{LDB}{long duration balloon}
    \acro{LED}{light emitting diode}
    \acro{LCS}{liquid cooling system}
    \acro{LC}{inductor and capacitor}
    \acro{LPE}{low-pass edge}
    \acro{MLR}{multilayer reflective}
    \acro{MAXIMA}{Millimeter~Anisotropy~eXperiment~IMaging~Array}
    \acro{NASA}{National Aeronautics and Space Administration}
    \acro{NDF}{neutral density filter}
    \acro{PCB}{printed circuit board}
    \acro{PE}{polyethylene}
    \acro{PME}{polarization modulation efficiency}
    \acro{PSF}{point spread function}
    \acro{PV}{pressure vessel}
    \acro{PWM}{pulse width modulation}
    \acro{RMS}{root mean square}
    \acro{SLR}{single layer reflective}
    \acro{SMB}{superconducting magnetic bearing}
    \acro{SQUID}{superconducting quantum interference device}
    \acro{SQL}{structured query language}
    \acro{STARS}{star tracking attitude reconstruction software}
    \acro{TES}{transition edge sensor}
    \acro{TDRSS}{tracking and data relay satellites}
   \acro{TM}{transformation matrix}
   \acro{UHMWPE}{ultra high molecular weight polyethylene}

\end{acronym}

\bibliographystyle{plain}
\bibliography{dwmbib}
\end{document}